\documentclass{Pos}

\title{The Stagnation of  Contemporary Stellar Astronomy}

\ShortTitle{The Stagnation of Contemporary Stellar Astronomy}

\author{\speaker{Petr \v{S}koda}\\
        Astronomical Institute, Academy of Sciences of the Czech Republic\\
        E-mail: \email{skoda@sunstel.asu.cas.cz}}

\abstract{
The stellar astronomy has always been considered  the fundamental source of
knowledge about the basic building blocks of the universe --- the stars. It has
proved correctness of many physical theories --- like e.g.  the idea of nuclear
fusion in stellar cores, the exchange of mass in interacting binaries or
models of stellar evolution towards white dwarfs or neutron stars.  Despite its well
acknowledged importance  it seems to be loosing its interestingness for
students, for telescope allocation committees at large observatories, as well as
for granting agencies.  In the domain of big telescopes it has been gradually
overtaken by the extra-galactic research and cosmology, surviving however at
smaller observatories and among most advanced amateur astronomers.

We try to analyse the main obstacles lowering the
efficiency of research in contemporary stellar astronomy. We will shortly
tackle several  problems induced by paradigmatic changes in handling the
extraordinary amount of data provided by current instruments as well as by
introduction of economical criteria  and  factory-like management into the
modern astronomy. 

Finally we speculate  the reasons of a marginal role of Virtual observatory  in
contemporary stellar research and give some ideas of possible improvements.
}

\FullConference{Accelerating the Rate of Astronomical Discovery, sps5\\
		August 11-14, 2009\\
		Rio de Janeiro, Brazil}

\begin{document}

\section{Introduction}

The technological potential of the current astronomical research is enormous.
The opening of almost all windows of electromagnetic spectrum facilitated by
advances in detector technology together with specialized astronomical
satellites as well as 8 to 10m class telescopes harnessed by extremely
sensitive instrumentation provides the astronomical community with
overwhelmingly massive amount of data about the very deep structure of our
universe as well as detailed multi-wavelength information about the billions
of even very distant objects. The most powerful GRIDs of computers as well as
clever infrastructure of Virtual Observatory are supposed to easily handle the
data avalanche and so amount of principal astronomical discoveries should
follow the technological Moore's law. But is it really so ?

Unfortunately, there is a danger of loosing the important dimension --- the
time --- from observations of the most powerful telescopes and their archives.
The astronomy of past was  based on long-term monitoring of selected objects
and most physical analysis was concerning their  time-dependent nature (e.g.
the search of periods, phase dependent changes of line profiles or changes of
intensity or spectra during the outbursts of novae or cataclysmic variables).
The largest amount of data coming from large facilities is currently only some
kind of snapshots of the Universe, being forced by the way the telescope time
is allocated  and by pressure on rapid publication of results.  All of this may
lead  to the paradoxical situation when the next generation of astronomers will
have from the then available data the feeling of the universe like something static
and unchangeable as it was in the ancient times before the Galileo's invention
of telescope.

\section{The Historical Importance of Stellar Astronomy}

\subsection{Breaking of Medieval Dogmas}  

The principal break in a philosophical concepts of Middle Age started by
observations of stellar astronomy. Before then the Universe was considered to
be static (Aristotle) and there was a generally accepted dogma postulating the
Sphere of Immutable Heavens is far away than the Sublunary spheres (filled with
fire, air, water and earth) where everything changing happens \cite{aristotle}.

The observation of Tycho's Supernova in 1572 was a nice opportunity to prove or
disprove this concept.  The observations of Tycho Brahe and Tadeas Hajek from
Hajek tried to estimate the distance of the Supernova.  As no parallax was
observed, the results clearly proved the supernova is much further away than
Moon thus breaking the idea of Immutable Heavens \cite{hajekall}.

\subsection{Physical Estimates from Variable Stars}

For enabling the first estimates of stellar physical quantities there were very
important discoveries of variable stars like Miras (omicron Cet by Fabricius
1596), Cepheids (delta Cep by Goodricke in 1784) as well as  justification of
nature of eclipsing binaries (the Algol by J. Goodricke in 1783)
\cite{firstvar}. Together with first confirmation of physical binaries (by W.
Herschel in 1803) \cite{binaries} the astronomers were given methods of
measuring the masses and radii of stars.

\subsection{The Role of Stellar Spectroscopy}

The important role in stellar research has played the spectroscopy The
discovery of spectral lines in solar spectrum (by Wollaston in 1802) and their
high resolution mapping  (by Fraunhofer in 1814) \cite{spectrum} together with
the development of laboratory spectroscopy (by Bunsen and Kirchhoff in 1859)
\cite{specanalysis} started the boom of stellar spectroscopy after the first
installation of spectroscope at the telescope (by Huggins arround 1860)
\cite{hugginsbib}.

The fundamental quantitative methodology for study of stellar dynamics was laid
after the first measurement of radial velocity of  Sirius (by Huggins in 1868)
using the theory of Doppler principle (postulated in 1842) \cite{hugginsbib}
\cite{huggins1}. 

An important moment for understanding the physical nature of stars presented
the method of stellar classification. The first trials of Sechhi in 1863
\cite{secchi} and advanced scheme introduced by  Pickering (in 1885) followed
the huge and tedious work of H. Draper and mainly A. Cannon \cite{cannon}  were
finally acknowledged by astronomical community in 1922 as an official  spectral
classification standards.

Another important step towards the understanding of the Universe 
was the discovery of relation Period---Luminosity of Cepheids in SMC
by H. Leavitt (in 1912) \cite{leavittorig}, which gave the astronomers the
tool allowing the measurement of large astronomical distances. 

The first important extragalactic measurements were obtained by Hubble (1923)
using the Cepheids to measure the distance of M31 and thus finishing the Great
Debate about its galactic nature \cite{hubbleorig}. In fact the currently most
frequently cited variable in extragalactic research --- the Red Shift --- is
just the logical combination of the techniques described earlier and thus a
result of stellar astronomy research.

\section{The Time Domain of Stellar Astronomy}

It is sometimes said that the astronomy is entering the time domain using the
network of robotic telescopes  to react properly at the alerts from satellites
--- mainly for hunting of gamma ray bursts events.  

But the stellar astronomy, in fact, had already entered the time domain at its
early origins.  Many physical ideas (e.g. the Roche lobe overflow scenario, or
theory of accretion disks, non radial pulsation modes, the  chemically peculiar
spots at surface of some stars) came  from the systematic spectroscopic
observations of variable stars on different time scales.  Deep knowledge  about
the interior of stars (and justification of theories) was obtained  from
observation of stellar  pulsations (Cepheids, RR Lyr, Miras) 

And current asteroseismology is even promising the detailed mapping of
internal chemical composition of stellar cores.  Although some of such a
research has been done by photometry, the crucial role here still plays the
time-resolved high resolution spectroscopy allowing the investigation of tiny
changes in spectral line profiles.

There are huge surveys, like VST, VISTA or LSST, planned gathering the time
series of observations \cite{surveys}. But all of them will give
only the photometric information. The sophisticated colour indices are expected
to yield the potential candidates for interesting classes of objects.  But the
problem is that there will be only candidates. The only proof of their real
physical nature will give only detailed spectroscopy.  Unfortunately, there are
no huge all-sky spectroscopic surveys planned with sufficient spectral coverage
and reasonably high resolution for stellar studies \cite{nospecsurveys}. 

There are, however  projects focused on time series observations of stellar
objects, however most of them are focused on the suspected extrasolar planet
and not on the star and its behaviour itself. Thus the selection of stars is
biased, because the formation of planets is expected mostly around later-type
solar analogues. 

Many interesting aspects of physics occurring in the hot early-type stars like
e.g. strong winds at O stars or  disks and circumstellar envelopes of Be and
B[e] stars are thus not getting a proper interest as they deserve --- mainly
their time evolution.

\section{Boom of Extragalactic Research}

As the sizes of telescopes and sensitivity of instrumentation became large
enough the interest of many astronomers switched naturally towards the objects
in distant Universe. The fascination by  AGNs and their X-ray  production or
the achievements of cosmology studying the most distant quasars and large-scale
structure and especially the hunt for dark  matter have changed not only the
image of contemporary astronomy among general public, but had strongly
influenced the goals and priorities of the astronomical research at all.

The long-run observing programs of many large telescopes allocates most of time
to extragalactic studies as well (typically to redshift surveys with integral
field units and multi-object spectrographs).  Most of the stellar projects just
get few nights and the telescope allocation committee's priority is put on
programs where one exposure will justify or rule-out some hypothesis.

So we have only a snapshots of the Universe enforced by large telescope time
allocation rules and general interests. The extragalactic bias is seen in
literature as well.

The publishers of main journals (and their peer reviewers) are accepting for
publication almost every  description of observation of an exotic distant
object, while they often require the publication of observation in stellar
astronomy to be accompanied by physical estimates and a model. But as the
gathering of proper material allowing the construction of a reasonable model
takes a long time and the theory of certain phenomena is still lacking behind
the observations, the amount of published articles in stellar astronomy has
decreasing in comparison to the extragalactic ones.

All these issues are responsible for the further  fostering of a role of
extragalactic research in future astronomical society as the interest of
majority of PhD. students is now focused mainly on  extragalactic research and
cosmology and the stellar astronomers are becoming the astronomical minority.

Similar concerns were expressed by A.Tokovinin at the conference about binary
stars in  Brno in June 2009 \cite{tokovinin} as the concept of a  Data Gap. He
had warned against the  noticeable gap in observations of binary stars
(especially missing time series of stellar spectra) occurring at the beginning
of 21st century to future astronomers.

\section{The Importance of Stellar Astronomy for Contemporary Extragalactic
Research}

The stellar research is, however,  crucial for understanding of the galactic
evolution as well as the structure of the whole universe.  

G. Bruzual has emphasized the key role of knowledge of stellar evolution in
understanding galaxies at the IAU General Assembly symposium S262  in Rio De
Janeiro:	

"There has been collected more photons from distant galaxies  than from nearby
stars but the  answers to distant galaxy problems will come from understanding
nearby stars !" 

The main issue is the construction of the stellar synthesis models used to
disentangle spectra of galaxies.  The problem of current  models is  connected
mainly with evolution of binary stars. During the post-AGB evolution the
thermally pulsating asymptotic giant branch stars contribute about 60\% of
K-band light \cite{reviewSS}, but about 30\% of them are binaries at Roche lobe
overflow  phase \cite{vanbeveren}. This is, however, usually ignored in most
models of stellar population synthesis.

For precise modelling we need to understand the detailed principles of many
phenomena observed on stars. Despite the progress in solution of many principal
questions of modern physics (e.g. the CNO and p-p cycles, structure of neutron
stars, sources of energy in black holes etc.) there are still open questions
of stellar astronomy, which have not yet been solved. Let's name some of them:
\begin{itemize}
\item Nature of binary star formation
\item The proper scenario of Post-AGB evolution in binaries
\item Proper physical explanation of origin and structure of disks around Be
and B[e] stars
\item The precise contents of primordial Lithium in different structures of
Universe --- it is a cosmological question which may be answered by  stellar
spectroscopy
\item The nature of stellar non-radial pulsations --- are they really
multi-periodic or  is it just a chaotic behaviour?
\end{itemize}

\section{Stellar Astronomy in Virtual Observatory}

The hope of astronomy facing enormous amount of data is focused to the Virtual
observatory allowing new types of pan-spectral research (e.g. easy building of
spectra energy distributions) and application of data mining methods on huge
surveys and databases of theoretical simulations.  But why the VO is not
widely used by the majority of stellar astronomers ?

The one of the main obstacles is the technological conservatism of  leading
stellar astronomers, who were accustomed to process every spectrum individually
from scratch and are very afraid of automatic processing of huge amount of
data. They do not trust the idea of VO and simply ignore it. But, what is worse,
they do not speak about the VO with young students, who might be interested
but are not informed. 

Thus most of the work in well-renowned stellar research groups at smaller
observatories and many universities has been still done using the
legacy applications (IRAF, MIDAS, IDL)  and custom (mostly
FORTRAN-coded) programs and scripts calling the different inhomogeneous data
converters from/to ASCII tables and simple plotting tools. 

But not only the conservatism plays its role here. There are people eager to
use the VO technology in stellar research, but they are paradoxically facing
the lack of stellar data not only in VO but in publicly available archives at
all.

Critical  situation is with stellar spectra archives. The data often play the
role of some kind of currency on the data exchange market and many astronomers
consider it to be their private property. So they are very reluctant in the
willingness of the  release of data even very long time after the publication. 

But the so called "data jealousy" is not the only reason. The most common reason
of lack of data archives of small observatories is connected with availability
of software engineers and data administrators capable of preparation and
maintenance of proper (VO-compatible) data archives.

There are even plans of artificial degradation of  published raw data in order
to prevent some kind of research by non-involved team --- especially for planet
hunting (e.g. high-pass filter had been used for KEPLER mission to prevent
the discovery of transiting extrasolar planet but allowing still the
asteroseismology studies \cite{kepler}, or the SOPHIE \cite{SOPHIE} archive of spectra is
missing exact time information for targets with a suspected extrasolar planets
\cite{prugniel}).

\section{The Role of Amateurs in Astronomy}

While the professional astronomers are attracted by extragalactic topics  under
the pressure of rapid publication of results (and partly to justify the
building of extremely large telescopes), there is a lot of very knowledgeable
amateurs understanding the role of their astronomical observations as a voluntary
work done for joy of knowledge.  

They are indispensable for continuation of stellar research. They are able to
concentrate on long-time monitoring of a large sample of stars as well as
prepare ingenious projects exploiting fully the capabilities of their modest
instrumentation \cite{hobbyast} (e.g. the Whole Earth Telescope, projects for
monitoring potential supernovae or transiting extrasolar planets and namely
large networks of monitoring eclipsing binaries and variable stars in general
--- mostly by CCD photometry) .The most advanced of them started building small
spectrographs and observe with them \cite{hobbyspec}.

However the limits of amateur observations are imposed by their typical
instrumentation --- the Peltier-cooled CCD photometry at 60cm telescopes or
their networks. Some of them are using commercial photo (still)  cameras or
even cameras with a film.  Despite their modest instrumentation they are the
primary source of long-term monitoring data about most variable stars and many
are regularly discovering new galactic novae  or even supernovae in distant
galaxies.

In contrast to this a lot of the  stellar spectroscopy has been often
accomplished at smaller and moderately large (1--3m) telescopes by students of
astronomy with their supervisors or smaller institutional  teams focused on
certain projects. 

The great potential of amateurs in their involvement in Virtual observatory is
still not fully appreciated.  They do not suffer of such a high level of
conservatism as professionals, are very eager to learn new technology and some
of them are prepared to offer their data in the VO --- just they have to be
informed about such a possibility.  The level of knowledge of VO among general
astronomical public is, unfortunately, still very low.

\section{Conclusions}

Stellar astronomy seems to be  now in stagnation period. The interest of
professional astronomers is turned towards very distant extragalactic objects.
The still existing professional stellar astronomy has been driven towards
producing only  snapshots  than  systematic monitoring.  The stars are
interesting for many astronomers only thanks to their planets, whereas most of
the physical problems of the stellar structure and stellar evolution as well as
nature of some phenomena remain still unsolved.  Unfortunately, the stellar
research seems  not to be very appealing for students either, despite the great
importance of stellar astronomy for understanding of galactic structure and
evolution.

On the other hand there is a lot of enthusiastic people among amateur
astronomers, who continue to monitor the stellar variability on the long time
scales and report outburst of cataclysmic variables, but they can hardly take
spectra of most interesting stars, mainly due to their limited capabilities.

The solution of the problems described might be the Virtual observatory,
allowing the integration of dispersed observations from different (even small)
teams.  The VO principles might attract the number of young people fascinated
by new software technologies.  Unfortunately, there is a lack of stellar
spectra in VO archives as well as little support of specific techniques
important in stellar astronomy in VO tools with respect to existing legacy
applications.  There is also very little VO awareness among stellar astronomers
and their students.  Thus the possible solution of the current stellar crisis
could be expressed by several goals:
\begin{itemize}
\item Convince people to put more  stellar spectra into VO archives
\item Circumvent the data jealousy by convincing people about advantage of openness 
\item Get resources needed for development of new  more versatile analysis tools with
	VO interface (or for adding  the VO interface to legacy applications)
\item Rise the level of VO awareness by systematic VO education or rather "VO evangelization" 
\end{itemize}
We can hopefully still preserve the image of the dynamic eternally changing
universe  for  our grandsons instead of hinting them the idea of the Immutable
Heavens again.

\end{document}